\documentclass[12pt]{article}
\usepackage{amsmath,amssymb,bm,epsfig}
\advance \topmargin by -\headheight
\advance \topmargin by -\headsep
     
\evensidemargin \oddsidemargin
\makeatletter
\renewcommand\section{\@startsection {section}{1}{\z@}%
                                   {-3.5ex \@plus -1ex \@minus -.2ex}%
                                   {2.3ex \@plus.2ex}%
                                   {\normalfont\large\bfseries}}
 
\renewcommand\subsection{\@startsection{subsection}{2}{\z@}%
                                   {-3.25ex\@plus -1ex \@minus -.2ex}%
                                   {1.5ex \@plus .2ex}%
       {\normalfont\normalsize\bfseries}}
\renewcommand{\subsubsection}{\@startsection{subsubsection}{3}{0mm}
   {0.75\baselineskip}%
   {0.5\baselineskip}%
   {\normalfont\normalsize\slshape\underline}}%
\makeatother
\renewcommand\d{\partial}
\newcommand\x{\mathbf{x}}
\newcommand\y{\mathbf{y}}

\newcommand\q{\mathbf{q}}
\renewcommand\j{\mathbf{j}}
\newcommand\dlr{\raisebox{0.1em}{$\stackrel{\scriptstyle\leftrightarrow}\partial$}}
\newcommand\+{\dagger}
\renewcommand\v{\mathbf{v}}
\newcommand\bfnab{\bm{\nabla}}
\newcommand\<{\langle}
\renewcommand\>{\rangle}
\newcommand\stru{\rule[-6mm]{0mm}{6mm}}
\newcommand\kF{k_{\mathrm{F}}}
\newcommand\vF{v_{\mathrm{F}}}
\newcommand\epsF{\epsilon_{\mathrm{F}}}
\renewcommand\L{\mathcal{L}}
\newcommand\conf{{\mathrm{c}}}

\begin{document}
\title{General coordinate invariance and conformal invariance in 
nonrelativistic physics: Unitary Fermi gas}
\author{D.~T.~Son and M.~Wingate\\
{\small\em Institute for Nuclear Theory,
University of Washington, Seattle, WA 98195-1550, USA}}

\date{September 2005}
\maketitle
\begin{abstract}
  We show that the Lagrangian for interacting nonrelativistic
  particles can be coupled to an external gauge field and metric tensor
  in a way that exhibits a nonrelativistic version of general
  coordinate invariance.  We explore the consequences of this
  invariance on the example of the degenerate Fermi gas at infinite
  scattering length, where conformal invariance also plays an
  important role.  We find the most general effective Lagrangian
  consistent with both general coordinate and conformal invariance to
  leading and next-to-leading orders in the momentum expansion.  At
  the leading order the Lagrangian contains one phenomenological
  constant and reproduces the results of the Thomas-Fermi theory and
  superfluid hydrodynamics.  At the next-to-leading order there are
  two additional constants.  We express various physical quantities
  through these constants.
\end{abstract}

\vspace{-15cm}
\begin{flushright}
INT-PUB 05-23
\end{flushright}

\newpage

\section{Motivation}

Symmetry is one of the most important ideas in modern physics.
Frequently, the identification of the symmetries of a physical system
is the key to understanding it.  In this paper we identify a new type
of symmetries of nonrelativistic physics, which are manifest when the
system is put in curved space, yet less evident in flat space.

The uncovering of these symmetries comes from our attempt to
understand the properties of the so-called ``unitary'' Fermi gas,
which is a gas of fermions which interact with each other through a
zero-range two-body potential tuned to infinite scattering length
$a=\infty$.  This system has been studied theoretically for some
time~\cite{Eagles,Leggett,Nozieres}, but only recently did it become
realized in experiments with trapped atoms using a Feshbach
resonance~\cite{Grimm,Jin,Ketterle,Thomas,Salomon}.  What is special
about this system the absence of any intrinsic parameter, except for
the number density.  Unfortunately, this same feature means that there
is no small coupling constant on which to build a reliable
perturbation theory.  It is therefore imperative that we find and
exploit all the symmetries of this system.

Despite the strong coupling, we are now certain that the ground state
of the unitary Fermi gas is a superfluid.  We also have, as our
starting point, at least two theoretical frameworks which allow one to
solve certain problems involving the unitary Fermi gas in a reliable
fashion.  These are the Thomas-Fermi theory and superfluid
hydrodynamics.  The two theories are very similar: both make no
attempt to solve the problem at the microscopic level; they are both
valid in the long-wavelength regime, when the characteristic
length scale of the problem is much larger than all microscopic scales
(including the healing length); and they need as an input at most one
parameter (which is denoted as $\xi$ in Eq.~(\ref{xi-def}) and
characterizes the zero-temperature equation of state.)  The total
energy and the oscillation frequencies of unitary Fermi gas in a
trap~\cite{BaranovPetrov,Stringari} can be computed using these
approaches.  The input parameter $\xi$ is not fixed within these
theories, but can either be measured experimentally or computed by
Monte-Carlo simulations.

However, until now it is not possible to systematically include
corrections to either the Thomas-Fermi theory or superfluid
hydrodynamics.  At weak coupling there is no fundamental obstacle to
compute these corrections, but our tools do not work at strong
coupling.  These corrections can be considerable if we want to study
the physics at length scales not much larger than the healing length.

The main, utilitarian motivation of this paper is to develop a
systematic way to go beyond the leading order in the long-wavelength
expansion.  This is done using an effective field theory
(EFT)~\cite{Weinberg:1996kr}.  The central idea of the EFT philosophy
is a power expansion over momentum (or inverse wavelength), instead of
the usual perturbative expansion over powers of the coupling constant.
This makes EFT particularly suitable for problems involving strongly
coupled systems in the long-wavelength limit.  Naturally we find that
at leading order the EFT reproduces the Thomas-Fermi theory and
superfluid hydrodynamics. However, in principle, the EFT can be
extended to arbitrary order in the expansion over momentum at the cost
of introducing an increasing number of free parameters.  The form of
the effective Lagrangian is constrained only by the symmetries of the
system under study.

Somewhat unexpectedly, putting all possible constraints on the
next-to-leading order (NLO) Lagrangian turns out to be a very
nontrivial problem in the case of the unitary Fermi gas.  In this
paper, this task is accomplished by temporarily putting the system in
curved space and exploiting the properties of general coordinate
invariance and conformal invariance.  Working in curved space
necessitates the use of certain mathematical tools normally employed
only in general relativity, but what we find at the end is quite
nontrivial and worth the effort.  In addition to one free parameter at
the leading order, one needs just two parameters to characterize the
NLO Lagrangian.  These parameters cannot be computed within the EFT,
but since the number of physical quantities that can be computed is
greater than three, the EFT has real predictive power.

We start the paper with a summary of some results that can be obtained
using the EFT (Sec.~\ref{sec:summary}).  Then in Sec.~\ref{sec:NRGCI}
we show that the Lagrangian for interacting nonrelativistic particles
can be coupled to background gauge fields and spatial metric in a way
that respects gauge invariance and the general coordinate invariance.
The general coordinate invariance is a gauge version of the
translation and Galilean transformations.  In Sec.~\ref{sec:conformal}
we study the conformal invariance of unitary Fermi gas.  We then
proceed, in Sec.~\ref{sec:Leff} and \ref{sec:NLO}, to find the most
general effective Lagrangian, to leading and next-to-leading orders,
that is consistent with these symmetries.  Some applications of the
EFT are considered in Sec.~\ref{sec:applications}.  We conclude with
Sec.~\ref{sec:conclusion}.

\section{Summary of results}
\label{sec:summary}

Here we give a brief summary of the result of this paper for the
unitary Fermi gas.  We do it here to give the reader a sense of the
power and the limitations of the effective field theory derived in the
paper.  We work in the unit system where $\hbar=1$.  We denote the
number density as $n$ and the fermion mass as $m$, from which we
define the Fermi momentum $\kF=(3\pi^2n)^{1/3}$, the Fermi energy
$\epsF=\kF^2/(2m)$, and the Fermi velocity $\vF=\kF/m$.  We introduce
the dimensionless parameter $\xi$, defined as the ratio of the energy
density of the unitary Fermi gas to the energy density of a free Fermi
gas at the same density,
\begin{equation}\label{xi-def}
  \epsilon = \xi \frac35 \epsF n.
\end{equation}

First we shall show that the dynamics of the gas is described, to
leading order, by the Lagrangian,
\begin{equation}
  \L_{\rm LO} = c_0 m^{3/2} \left[\mu-V(t,\x)-\dot\varphi-
   \frac{(\nabla\varphi)^2}{2m}\right]^{5/2},
\end{equation}
where $\varphi$ is the phase of the condensate, $\mu$ is the chemical
potential, $V(t,\x)$ is the trapping potential, and $c_0$ is a
dimensionless parameter.  The leading-order Lagrangian captures the
same information as the Thomas-Fermi theory and superfluid
hydrodynamics.  The coefficient $c_0$ is directly related to the parameter
$\xi$, defined in Eq.~(\ref{xi-def}), by
\begin{equation}
  c_0 = \frac{2^{5/2}}{15\pi^2\xi^{3/2}}\,,
\end{equation}
and has to be determined microscopically.

To the next-to-leading order in the low-momentum expansion, the
effective Lagrangian has the form
\begin{equation}\label{summary:LNLO}
  \L = c_0 m^{3/2} X^{5/2} 
  + c_1 m^{1/2} \frac{(\bfnab X)^2}{\sqrt X} +
  \frac{c_2}{\sqrt m} [ (\nabla^2\varphi)^2 - 9m \nabla^2 A_0] \sqrt X\,.
\end{equation}
where we have introduced the shorthand notation
\begin{equation}
  X = \mu - V(t,\x) - \dot\varphi - \frac{(\nabla\varphi)^2}{2m}\,,
\end{equation}
and $c_1$ and $c_2$ are two additional dimensionless parameters that
appear at the NLO level.

Many quantities can be computed using the effective Lagrangian.  Let
us consider infinite system first ($V=0$).  The dispersion relation
for low-energy phonons is
\begin{equation}\label{sum:dispersion}
  \omega(q) =  c_s q\left[1
   -\pi^2\sqrt{2\xi}\left(c_1+\frac32 c_2\right)\frac{q^2}{\kF^2}\right]
  + O(q^5\ln q), \qquad c_s = \sqrt{\frac\xi3}\, \vF\,.
\end{equation}
The static density and transverse response functions are
\begin{subequations}\label{sum:response}
\begin{align}
  \chi(q) &= - \frac{m\kF}{\pi^2\xi} \left[
    1 + 2\pi^2\sqrt{2\xi}\left(c_1 - \frac92 c_2\right) 
    \frac{q^2}{\kF^2}\right] + O(q^4\ln q),\\
  \chi^T(q) &=  -9c_2 \sqrt{\frac\xi2}\, \vF q^2 + O(q^4\ln q).
\end{align}
\end{subequations}
The predictive power of the EFT is illustrated by a relation between
the nonlinear $q^3$ term in the phonon
dispersion~(\ref{sum:dispersion}) and the static response functions,
\begin{equation}
  \omega = c_s q \left( 1 - \frac{\chi(q)-\chi(0)}{2\chi(0)} 
     + \frac{4\pi^2}3 \frac{\chi^T(q)}{\vF\kF^2}\right) .
\end{equation}

In addition, we also find the ground-state energy of a system of $N$
particles in a harmonic trap to be
\begin{equation}
  E = \frac{\sqrt\xi}4 \bar\omega (3N)^{4/3} -
  \sqrt2\, \pi^2\xi \left(c_1 - \frac92c_2\right)
  \frac{\omega_1^2+\omega_2^2+\omega_3^2}{\bar\omega}(3N)^{2/3}
  + O(N^{5/9}),
\end{equation}
where $\omega_i$, $i=1,2,3$ are the trap frequencies,
$\bar\omega\equiv(\omega_1\omega_2\omega_3)^{1/3}$.

\section{Nonrelativistic general coordinate invariance}
\label{sec:NRGCI}

\subsection{Statement of general coordinate invariance}

Let us first consider a system of noninteracting particles (fermions
or bosons).  Instead of letting particles move freely in flat space,
we will put the system in an external gauge field
\begin{equation}
   A_0 = A_0(t,\x), \quad  A_i = A_i(t,\x), \qquad i=1,2,3.
\end{equation}
and a curved three-dimensional manifold with the spatial line element
\begin{equation}\label{line-element}
  ds^2 = g_{ij}(t,\x)\, dx^i\, dx^j .
\end{equation}
As indicated in Eq.~(\ref{line-element}), the metric on the manifold
is, in general, time-dependent, so the manifold can change shape with
time.  The spatial indices $i,j,\ldots$ will be raised and lowered by
the metric $g_{ij}$ and its inverse $g^{ij}$.  The determinant of
$g_{ij}$ is denoted as $g$.  The action of the system is
\begin{equation}\label{free-L}
  S = \int\!dt\,d\x\, \sqrt g\, \L = 
  \int\!dt\,d\x\, \sqrt{g}\left[\frac i2 \psi^\+ \dlr_t\psi
  - A_0\psi^\+\psi 
  - \frac{g^{ij}}{2m}(\d_i\psi^\+-iA_i\psi^\+)(\d_j\psi+iA_j\psi)\right].
\end{equation}
Here $\psi^\+\dlr_t\psi \equiv \psi^\+\d_t\psi-\d_t\psi^\+\psi$.  Note
that $A_0$, $A_i$, and $g_{ij}$ appear as external parameters of the
Lagrangian; they have no dynamics of their own.  The most common
situation is when the system is put in a trapping potential $V$.  This
corresponds to $A_0=V$, $A_i=0$, $g_{ij}=\delta_{ij}$.  Another case,
not considered in this paper, is that of rotating systems where $A_i$
are not zero.  Although space is flat in all practical applications,
keeping $g_{ij}$ arbitrary in intermediate steps allow one to derive
nontrivial results, as we shall see.

We have written the time-derivative term in Eq.~(\ref{free-L}) in the
symmetric form $(i/2)\psi^\+\dlr_t\psi$.  In flat space this is
equivalent to the asymmetric form $i\psi^\+\d_t\psi$; the two differ
from each other by a full time derivative.  This continues to be true
if the metric is time-independent.  For a general time-dependent
metric, however, the two forms are not equivalent.  As at the end we
shall put the metric to flat, the choice between the two forms is a
matter of convenience.  We will soon find that the symmetric form
in~(\ref{free-L}) is much more advantageous.

The action~(\ref{free-L}) is invariant under gauge transformations,
\begin{equation}\label{gauge-tr}
  \psi \to \psi' = e^{i\alpha}\psi,\quad
  A_0 \to A_0' = A_0 - \dot\alpha,\quad
  A_i \to A_i' = A_i - \d_i\alpha,
\end{equation}
where $\dot\alpha\equiv\d_t\alpha$.  It is also invariant under a change
of coordinate on the 3D manifold,
\begin{equation}\label{3d-gci}
\begin{array}{ll}
  x^i \to x^{i'} = x^{i'}(x^i), & \\
  \psi(t,\x) \to \psi(t,\x') = \psi(t,\x), &
  A_0(t,\x)\to A_0'(t,\x') = A_0 (t,\x), \\
  g_{ij}(t,\x) \to g_{i'j'}(t,\x') = \dfrac{\d x^i}{\d x^{i'}}
    \dfrac{\d x^j}{\d x^{j'}} g_{ij}(t,\x), \qquad&
    A_i(t,\x) \to A_{i'} (t,\x') = \dfrac{\d x^i}{\d x^{i'}}
    A_i(t,\x).
\end{array}
\end{equation}
The infinitesimal version of (\ref{gauge-tr}) and (\ref{3d-gci}) is
\begin{equation}\label{static-gci}
\begin{array}{ll}
  \delta\psi = i\alpha\psi - \xi^k\d_k\psi, 
  & \delta A_0 = -\dot\alpha -
    \xi^k\d_kA_0, \\
  \delta g_{ij} = -\xi^k \d_k g_{ij} - g_{ik}\d_j \xi^k -
     g_{kj}\d_i \xi^k, \quad &
  \delta A_i = -\d_i\alpha - \xi^k\d_k A_i - A_k \d_i\xi^k .
  \end{array}
\end{equation}
We note that while the gauge parameter $\alpha$ in
Eq.~(\ref{static-gci}) can be a function of both time and space, so
far we have considered the gauge parameter $\xi^i$ (which is the
coordinate shift, $x^i\to x^i+\xi^i$) to be time-independent.  In
fact, if one attempts to make $\xi^i$ time-dependent, one finds that
the action is not invariant, but acquires a change proportional to
the time derivative of $\xi^i$,
\begin{equation}
  \delta S = -\frac i2\!\int\!dt\,d\x\, \dot\xi^k \psi^\+\dlr_k\psi .
\end{equation}
However, the invariance of the action can be restored by a small
modification of the transformation laws~(\ref{static-gci}).  The
modified transformations read
\begin{subequations}\label{nonrel-gci}
\begin{align}
  \delta\psi &= i\alpha\psi -\xi^k\d_k\psi, \\
  \delta A_0 &= -\dot\alpha-\xi^k\d_k A_0 - A_k \dot\xi^k ,
  \label{nonrel-gci-A0}\\
  \delta A_i &= -\d_i\alpha-\xi^k\d_k A_i - A_k\d_i\xi^k + mg_{ik}\dot\xi^k,
  \label{nonrel-gci-Ai}\\
  \delta g_{ij} &= -\xi^k\d_k g_{ij} -g_{ik}\d_j\xi^k -
  g_{kj}\d_i\xi^k .
\end{align}
\end{subequations}
It is straightforward to verify that terms proportional to $\dot\xi^k$
now cancel in $\delta S$.  (Had we written the time-derivative term in
the asymmetric form $i\psi^\+\d_t\psi$ this would not be true.)  We
shall call (\ref{nonrel-gci}) the \emph{nonrelativistic general
coordinate transformations}, or simply general coordinate
transformations.  The action is said to be general coordinate
invariant.

Let us discuss the relationship between the
transformations~(\ref{nonrel-gci}) with the more familiar Galilean
transformations.  In contrast to~(\ref{nonrel-gci}), the Galilean
transformations are defined in flat space, $g_{ij}=\delta_{ij}$.  They
can be thought of as a special class of~(\ref{nonrel-gci}) with a
particular choice of $\alpha$ and $\xi^i$:
\begin{equation}\label{alphaxiGal}
  \alpha = mv^i x^i, \qquad \xi^i = v^i t .
\end{equation}
This leaves the flat metric unchanged while transforming other fields,
\begin{equation}\label{Galilean}
\begin{array}{l}
   \psi(t,\x) \to \psi'(t,\x) = e^{im\v\cdot \x} \psi(t,\x-\v t),  \\
   A_0(t,\x) \to A_0'(t,\x) = A_0(t,\x-\v t) - v^k A_k(t, \x-\v t),\\
   A_i(t,\x) \to A_i'(t,\x) = A_i(t,\x-\v t).
\end{array}
\end{equation}
Note that in Eqs.~(\ref{Galilean}) terms proportional to $v^2$ and
higher orders of $v$ have been dropped.  Thus any theory with general
coordinate symmetry is automatically Galilean invariant when
restricted to flat space.%
The transformations~(\ref{nonrel-gci}) can be thought of as a gauge
version of the translation and Galilean transformations.  A similar
``gauging'' of nonrelativistic spatial translations has been
considered previously~\cite{Lukierski:2000pq}.  A subclass
of~(\ref{nonrel-gci}) which does not change the flat metric has been
considered in Ref.~\cite{Jackiw:1992fg}.

So far we have considered the noninteracting theory.  It is easy to
introduce interaction into the theory in a way that respects general
coordinate invariance by introducing auxiliary fields.  For example,
the theory with the following Lagrangian,
\begin{equation}\label{int-L}
  \L = \frac i2 \psi^\+ \dlr_t\psi
  - A_0\psi^\+\psi 
  - \frac{g^{ij}}{2m}(\d_i\psi^\+-iA_i\psi^\+)(\d_j\psi+iA_j\psi)
  + q_0\psi^\+\psi\sigma - \frac12 g^{ij}\d_i\sigma\d_j\sigma
  - \frac {\sigma^2}{2r_0^2}\,,
\end{equation}
is consistent with general coordinate invariance, as one can make
$\sigma$ transform as
\begin{equation}
  \delta\sigma = -\xi^k \d_k \sigma .
\end{equation}
In flat space (\ref{int-L}) describes a system of particles
interacting though a Yukawa two-body potential,
\begin{equation}
   V(r) = -\frac{q_0^2}{4\pi r} e^{-r/r_0}.
\end{equation}
We thus have an example of an interacting theory where coupling to an
external gauge field and a metric can be introduced in a way
consistent with general coordinate invariance.

We have not demonstrated that general coordinate invariance can be
achieved for \emph{any} two-body potential.  However, our foremost
interest is in systems with universal behavior, where only the
scattering length $a$ is important, not the details of the two-body
potential.  For this purpose, the Lagrangian~(\ref{int-L}) is
sufficient.  In particular, infinite scattering length is achieved by
fine-tuning $q_0$ and $r_0$ so that
\begin{equation}\label{fine-tuning}
  mq_0^2r_0 = 21.1...
\end{equation}
when the first bound state appears at threshold.  To achieve
universality we also need to keep $r_0$ smaller than any physical
length scale.  The universality of the fermionic system with infinite
scattering length means that one can choose $q_0$ and $r_0$
arbitrarily (keeping $r_0$ small), and the physics does not depends on
our choice if Eq.~(\ref{fine-tuning}) is satisfied.  One can even make
$q_0$ and $r_0$ time-dependent and nothing will depend on their
values, if the fine-tuning condition is kept at all time.  This remark
will be important in our subsequent discussion.

We have found the transformations~(\ref{nonrel-gci}) by trial and
error.  An alternative way to arrive to these transformations is by
taking the nonrelativistic limit of the relativistic general
coordinate invariance.  The second way is instructive and will be
useful for us when we try to write down an effective Lagrangian
consistent with general coordinate invariance.  We will describe it in
the next Section.

\subsection{The relativistic root of nonrelativistic general
coordinate invariance}

Consider a relativistic field theory of one free complex scalar field
$\Psi$ in a external four-dimensional metric $g_{\mu\nu}$.  We use the
$({-}{+}{+}{+})$ metric signature with $x^\mu=(ct,\,\x)$ and
$ds^2=g_{\mu\nu}dx^\mu dx^\nu$.  We keep track of the speed of light
$c$ in order to send $c\to\infty$ in the nonrelativistic limit.  The
action,
\begin{equation}\label{relat-S}
  S = -\int\!d^4x\, \sqrt{-g}\, \left( g^{\mu\nu}\d_\mu\Psi^*\d_\nu\Psi 
       +m^2c^2\Psi^*\Psi\right),
\end{equation}
is invariant under the infinitesimal general coordinate transformation
\begin{subequations}\label{relat-gci}
\begin{align}
  \delta\Psi &= - \xi^\lambda\d_\lambda \Psi,\\
  \delta g_{\mu\nu} &= -\xi^\lambda\d_\lambda g_{\mu\nu} 
    -g_{\lambda\nu}\d_\mu\xi^\lambda - g_{\mu\lambda}\d_\nu\xi^\lambda .
\end{align}
\end{subequations}

In the nonrelativistic limit we take $c\to\infty$.  We assume that in
this limit $\Psi$ has a fast phase rotation in time solely due to its
rest mass, which can be factored out to yield the nonrelativistic
field $\psi$,
\begin{equation}\label{Psipsi}
  \Psi = e^{-imcx^0}\frac\psi{\sqrt{2mc}} 
  = e^{-imc^2t}\frac\psi{\sqrt{2mc}}\,.
\end{equation}
The metric is taken to have the $(00)$ component close to $-1$ and small
$(0i)$ components,
\begin{subequations}\label{GAg}
\begin{equation}
   g_{\mu\nu} = \left( \begin{array}{ccc}
   -1 - \dfrac{2A_0}{mc^2} &
   &-\dfrac{A_i}{mc}\rule[-6mm]{0mm}{6mm}
  \\
    -\dfrac{A_i}{mc} & & g_{ij} \end{array}\right),
\end{equation}
where $A_0$, $A_i$, and $g_{ij}$ are kept fixed in the limit
$c\to\infty$.  The inverse metric is
\begin{equation}\label{Ginv}
  g^{\mu\nu} = \left( \begin{array}{ccc}
   -1 + \dfrac{2A_0}{mc^2} 
     +\dfrac{A^iA_i}{m^2c^2}+O(c^{-4})&
   &-\dfrac{A^i}{mc} + O(c^{-3})\rule[-6mm]{0mm}{6mm}\\
   -\dfrac{A^i}{mc} + O(c^{-3}) & & 
   g^{ij} + O(c^{-2})
  \end{array}\right),
\end{equation}
\end{subequations}
where $A^i\equiv g^{ij}A_j$.  Substituting Eqs.~(\ref{Psipsi}) and
(\ref{Ginv}) into the action~(\ref{relat-S}), one reproduces
(\ref{free-L}) after discarding terms containing negative powers of
$c$.  Note that the $g_{00}$ and $g_{0i}$ components of the 4D metric
tensor now appear as the gauge field in the nonrelativistic action, in
a manner reminiscent of the Kaluza-Klein
mechanism~\cite{Appelquist:1987nr}.

Taking the coordinate shift $\xi^\mu$ of the form
\begin{equation}
  \xi^\mu = \left( \frac\alpha{mc}\,,\ \xi^i \right),
\end{equation}
where $\alpha$ is fixed in the limit $c\to\infty$,
Eqs.~(\ref{relat-gci}) reduce to Eqs.~(\ref{nonrel-gci}).  Thus we can
understand (\ref{nonrel-gci}) as a nonrelativistic limit of the
relativistic general coordinate invariance.

\subsection{General coordinate invariance and conservation laws}

We now show that the general coordinate invariance implies
conservation of particle number and momentum.  We also will find that
in theories with general coordinate invariance the number current
coincides (up to a factor of $m$) with the momentum density.

The connected part of the generating functional $W$ is formally
defined as a path integral
\begin{equation}
  e^{iW[A_0,A_i,g_{ij}]} = \int\!D\psi^\+ D\psi\, 
  e^{iS[\psi,\psi^\+,A_0,A_i,g_{ij}]} \,.
\end{equation}
The correlation functions of operators coupled to the sources can be
found by differentiating $W$ with respect to its arguments.
General coordinate invariance implies, to linear order in $\alpha$ and
$\xi^i$,
\begin{equation}\label{Wdelta}
  W[A_0+\delta A_0, A_i+\delta A_i, g_{ij}+\delta g_{ij}]
  = W[A_0,A_i,g_{ij}],
\end{equation}
where $\delta A_0$, $\delta A_i$, and $\delta g_{ij}$ are defined as
in Eqs.~(\ref{nonrel-gci}).  Expanding the left-hand side of 
Eq.~(\ref{Wdelta}) to linear order in $\alpha$ and $\xi^i$, one finds
\begin{subequations}\label{conservation}
\begin{align}
   \d_t n + \d_i j^i &= 0,\\
   \d_t T_{0k} + \d_i T^i_k &= 0,
\end{align}
\end{subequations}
where
\begin{equation}
\begin{array}{ll}
  n = - \dfrac{\delta W}{\delta A_0}\,, & 
     T_{0k} = -m g_{ik}\dfrac{\delta W}{\delta A_i}
     + A_k\dfrac{\delta W}{\delta A_0}\,,\stru\\
  j^i = - \dfrac{\delta W}{\delta A_i}\,, \qquad &
  T^i_k = 2g_{kj}\dfrac{\delta W}{\delta g_{ij}}-\delta^i_k W 
            + A_k\dfrac{\delta W}{\delta A_i}\,.
\end{array}
\end{equation}
Eqs.~(\ref{conservation}) can be interpreted as conservation of
particle number and momentum.  Here $n$ is the number density, $j^i$
is the number flux, $T_{0i}$ is the momentum density and $T^i_k$ is
the stress tensor.  We also notice that in flat space
and in the absence of the vector potential
the momentum density is proportional to the number current,
\begin{equation}\label{GWW-cond}
  T_{0i} = m j^i \quad\textrm{when}~A_i =0,\ g_{ij}=\delta_{ij}.
\end{equation}
This is the condition previously used by Greiter, Wilczek, and
Witten~\cite{Greiter:1989qb} to find the effective field theory of
superfluids.  Any theory which is general coordinate invariant
automatically satisfies (\ref{GWW-cond}).

\section{Scale and conformal symmetries}
\label{sec:conformal}

The Lagrangian of noninteracting particles~(\ref{free-L}) exhibits
scale invariance, which is an overall rescaling of time and fields,
\begin{equation}
\begin{array}{ll}
  t \to t' = \lambda^{-1} t , & \\
  \psi(t,\x) \to \psi'(t',\x) = \lambda^{3/4}\psi(t,\x),\qquad
     & A_0(t) \to A'_0(t') = \lambda A_0(t), \\
  g_{ij}(t) \to g'_{ij}(t') = \lambda^{-1} g_{ij}(t), &
     A_i(t) \to A'_i(t') = A_i(t) .
\end{array}
\end{equation}
It is useful to introduce the notion of \emph{scaling dimension}.  An
operator $O$ is said to have scaling dimension $\Delta_O$ (in
shorthand notation, $[O]=\Delta_O$) if under scale transformation in
transforms as
\begin{equation}
  O(t, \x) \to O'(\lambda^{-1}t, \x) = \lambda^{\Delta_O} O(t,\x).
\end{equation}
Note that the scaling dimension of a product is the sum of scaling
dimensions: $[O_1O_2]=[O_1]+[O_2]$.  Also, one can assign scaling
dimensions to space and time derivatives, $[\d_i]=0$ and $[\d_t]=1$,
so $[\d_i O]=[\d_i\d_j O]=[O]$, while $[\d_t O]=[O]+1$.  In this
language,
\begin{equation}\label{scaling}
  [t] = -1, \quad [x^i]= 0,\quad
  [\psi] = \frac34\,,
  \quad [A_0] = 1,\quad [A_i] = 0,\quad [g_{ij}] = -1.
\end{equation}
and $[\L]=\frac52$, so $S$ is invariant.\footnote{The zero scaling
dimension of $x$ and $A_i$ in Eq.~(\ref{scaling}) may seem somewhat
unnatural, but that is because we have assigned $g_{ij}$ a scaling
dimension of $-1$.}

In fact, the action is invariant under a larger set of
transformations, which involves arbitrary reparametrization of time
$t\to t'=t'(t)$, simultaneous with rescaling of $A_0$ and $g_{ij}$,
\begin{equation}\label{conf-large}
\begin{array}{ll}
  \psi(t) \to \psi'(t') = \left(
     \dfrac{dt}{dt'}\right)^{3/4}\psi(t),\stru\qquad
  & A_0(t) \to A'_0(t') = \left(\dfrac{dt}{dt'}\right)A_0(t), \\
  g_{ij}(t) \to g'_{ij}(t') = \left(\dfrac{dt'}{dt}
     \right) g_{ij}(t), &
  A_i(t) \to A'_i(t') = A_i(t).\\
\end{array}
\end{equation}
The infinitesimal version of~(\ref{conf-large}) is
\begin{equation}\label{conf-inf}
\begin{array}{ll}
  t \to t' = t +\beta, & \\
  \delta\psi = -\beta\dot\psi - \frac34 \dot\beta\psi , 
    & \delta A_0 = -\beta\dot A_0 - \dot\beta A_0, \\
  \delta g_{ij} = -\beta \dot g_{ij} +\dot\beta g_{ij}, \qquad&
   \delta A_i = -\beta\dot A_i .
\end{array}
\end{equation}
We shall call this transformation the \emph{conformal
transformation}.\footnote{In the
literature~\cite{Hagen:1972pd,Mehen:1999nd} the term ``conformal
transformation'' was used for a particular a combination
of~(\ref{conf-inf}) and the general coordinate
transformation~(\ref{nonrel-gci}) with $\beta=-Ct^2$, $\xi^k=-Ctx^k$
and $\alpha=-\frac12mC\x^2$, where $C$ is a constant.  This
combination leaves the trivial background $A_0=A_i=0$,
$g_{ij}=\delta_{ij}$ unchanged.}  The scale
transformation~(\ref{scaling}) is a particular case of conformal
transformations with $\beta$ being a linear function of time:
$\beta=bt$.

In general, if a field $O$ transforms under conformal transformation
as
\begin{equation}\label{conf-O}
  \delta O = - \beta\dot O - \Delta_O\dot\beta O,
\end{equation}
then $O$ is said to have conformal dimension $\Delta_O$,
$[O]_\conf=\Delta_O$.  Note that we still have $[O_1
O_2]_\conf=[O_1]_\conf+[O_2]_\conf$, and $[\d_i O]_\conf=[O]_\conf$.
However, we cannot assign a conformal dimension to the time derivative
$\d_t$.  In fact, if $O$ transforms as in Eq.~(\ref{conf-O}), then
in general
$\dot O$ does not have a well-defined conformal dimension at all,
\begin{equation}
  \delta\dot O = -\beta\ddot O - (\Delta_O+1)\dot\beta \dot O -
      \Delta_O \ddot\beta O .
\end{equation}
The extra term proportional to $\ddot\beta$ prevents us from assigning
the conformal dimension $\Delta_O+1$ to $\dot O$, unless $\Delta_O=0$.  
(This term is
absent for scale transformations where $\beta$ is linear in $t$.)
Therefore while an operator with a conformal dimension has a scaling
dimension, the reverse is not true in general.

Note that the Lagrangian~(\ref{free-L}) contains a term with a time
derivative but still have a well-defined conformal dimension of
$\frac52$, since the $\ddot\beta$ terms cancel out between the
variations of $\psi^\+\d_t\psi$ and $\d_t\psi^\+\psi$.  Again we see
that the symmetric from of the time-derivative term is preferable.

The conformal symmetry is generally broken when one turns on an
interaction.  We now show that conformal invariance becomes exact
again in the limit of infinite scattering length.  For that we can use
the model with Yukawa interaction~(\ref{int-L}).  If we take $\sigma$
to transform the same way as $\psi$,
\begin{equation}
  \sigma(t) \to \sigma'(t')=\left(\frac{dt}{dt'}\right)^{3/4}\sigma(t),
\end{equation}
then after substitution, we see that two terms in the action,
$q_0\psi^\+\psi\sigma$ and $-\sigma^2/(2r_0^2)$, change under the
conformal transformation.  The effect is that the constants $q_0$ and
$r_0$ are replaced by new time-dependent constants
\begin{equation}
  q_0 \to q_0\left(\frac{dt'}{dt}\right)^{1/4}, \qquad 
  r_0 \to r_0 \left(\frac{dt}{dt'}\right)^{1/2} .
\end{equation}
However, we recall that the condition of infinite scattering length
corresponds to $mq_0^2r_0$ equal the critical
value~(\ref{fine-tuning}).  We can see that this condition is not
violated under the conformal transformation.  Thus we can make use of
the freedom in choosing $q_0$ and $r_0$ to transform it back to the
original constant $q_0$ and $r_0$, without changing the generating
functional $W$.  Therefore, the generating functional of a system
tuned to infinite scattering length is conformally invariant.  This
property of unitary Fermi gas will be used in Sec.~\ref{sec:NLO}.

Finally, we note that in contrast to the general coordinate
transformations, the scale and conformal transformations described
above are not descended from the relativistic counterparts.

\section{The superfluid effective theory}
\label{sec:Leff}

\subsection{Preliminaries}

It is now well established that Fermi gas remain superfluid in the
whole range of the parameter $\kF a$ from the BEC to the BCS regimes.
Superfluids provide the simplest nontrivial application of general
coordinate invariance.  At long distance the only physical degree of
freedom is the phase of the condensate.  We shall therefore construct
an effective field theory of the this phase.

For a system consisting of bosons, the field is defined as the phase
of the microscopic field,
\begin{equation}
  \psi = |\psi| e^{-i\theta}.
\end{equation}
For fermions, we define $\theta$ as half of the phase of the Cooper
pair,
\begin{equation}
  \<\psi\psi\>  = |\<\psi\psi\>| e^{-2i\theta}.
\end{equation}

Frequently, one describes a system at finite chemical potential $\mu$
by including the term $\mu\psi^\+\psi$ into the Lagrangian.  This
approach is not suitable for our purposes since this term breaks
general coordinate, scale and conformal symmetries.  Instead we shall
write the Lagrangian without the $\mu\psi^\+\psi$ term.  The chemical
potential enters the problem through the choice of the ground state.
In particular, the superfluid ground state of a system at chemical
potential $\mu$ corresponds to $\theta=\mu t$.  We will frequently
expand
\begin{equation}\label{thetavarphi}
  \theta = \mu t - \varphi,
\end{equation}
where fluctuations of $\varphi$ around zero correspond to phonon
excitations.  The breaking of symmetries is now \emph{spontaneous},
i.e., it comes from the choice of the ground
state~(\ref{thetavarphi}).

Let us first discuss the theory in free flat space ($A_0=A_i=0$,
$g_{ij}=\delta_{ij}$).  The effective field theory is formulated as a
Lagrangian $L_{\rm eff}(\theta)$.  This Lagrangian has a shift
symmetry $\theta\to\theta+\textrm{const}$, due to the global symmetry
with respect to phase rotation of $\psi$.  Due to this symmetry,
$\theta$ does not appear without derivative.  So when we expand the
Lagrangian in power series over derivatives and over $\theta$, we
encounter in each term at least as many derivatives as $\theta$.
Consider a few representative term in this expansion:
\begin{equation}\label{L-symb-exp}
  \L_{\rm eff}(\theta) \sim (\d\theta)^2 + (\d\theta)^4 
       + (\d\d\theta)^2 + \cdots
\end{equation}

The most naive power counting scheme is to consider $\theta\sim1$ and
each derivative as $O(p)$.  In this scheme, the first term
in~(\ref{L-symb-exp}) is $O(p^2)$, while the last two terms are
$O(p^4)$.  However, the power counting scheme that is most useful is
to count $\d_t\theta$ and $\d_i\theta$ as $O(1)$, and additional time
and space derivatives as $O(p)$.  This is possible since, as we have
seen, $\theta$ always appears with at least one derivative.  In this
second power counting scheme, the first two terms
in~(\ref{L-symb-exp}) are of order 1 while the last term is $O(p^2)$.
A generic term in the which contains $N[\d_t]$ time derivatives,
$N[\d_i]$ spatial derivatives, and the $N[\theta]$ power of the field
will be counted as 
\begin{equation}
  \d_t^{N[\d_t]} \d_i^{N[\d_i]} \theta^{N[\theta]} \sim 
   p^{N[\d_t]+N[\d_i]-N[\theta]} .
\end{equation}

In the second power counting scheme, the leading order effective
Lagrangian sums up all terms where the power of $\theta$ is equal to
the total number of space and time derivatives.  Thus it is a function
of first derivatives of $\theta$,
\begin{equation}
  \L_{\rm LO} = L(\dot\theta,\d_i\theta),
\end{equation}
which may contain, upon series expansion, terms with any power of
$\d\varphi$.

So far we have not discussed the background fields $A_0$, $A_i$, and
$g_{ij}$.  Since $A_0$ and $A_i$ appears with in covariant
derivatives, it is natural to assume that all fields scale as $O(1)$
and their first derivatives as $O(p)$.  In fact, in a trap where one
has to treat the trapping potential $A_0$ as a quantity of order 1, the
second power counting scheme with $\d\theta\sim1$ is the only
consistent one.

\subsection{Symmetry requirements on the effective Lagrangian}

The effective theory should inherit the general coordinate invariance
of the microscopic theory.  This means that the effective action is
invariant under the general coordinate transformations.  Since
\begin{equation}
  \delta \sqrt g = \frac12\sqrt g\, g^{ij}\delta g_{ij} = 
  -\xi^k \d_k \sqrt g - \sqrt g\, \d_k \xi^k ,
\end{equation}
we need the Lagrangian density to transform as
\begin{equation}\label{deltaLscalar}
  \delta \L = -\xi^k \d_k \L ,
\end{equation}
so that the action is invariant
\begin{equation}
  \delta S = \int\!dt\,d\x\, \left(-\xi^k\L \d_k\sqrt g\, 
  - \sqrt g\,\L \d_k\xi^k - \sqrt g\,\xi^k \d_k \L\right) =
  -\!\int\!dt\,d\x\, \d_k(\sqrt g\, \xi^k \L) = 0.
\end{equation}
Equation~(\ref{deltaLscalar}) means that $\L$ has to transform as a
scalar under general coordinate transformations.

We shall assume that the microscopic theory is invariant under time
reversal, which means that the effective theory is invariant under
\begin{equation}
  t\to -t, \qquad \theta \to -\theta .
\end{equation}
This time reversal invariance has an important consequences for the
momentum expansion of the EFT.  This invariance implies that in each
term, the difference between of the number of time derivatives and the
number of $\theta$ is even,
\begin{equation}
  N[\d_t] - N[\theta] = \textrm{even} .
\end{equation}
Due to rotational symmetry and parity (which forbids
$\epsilon^{ijk}$), the number of spatial derivatives have to be even,
since the spatial indices have to contract pairwise.  Therefore we
have
\begin{equation}
  N[\d_t] + N[\d_i] - N[\theta] = \textrm{even}, 
\end{equation}
which means that the terms in the effective Lagrangian are
proportional to even powers of $p$.  Therefore the next-to-leading
order terms in the Lagrangian are $O(p^2)$, the
next-to-next-to-leading terms are $O(p^4)$, etc.

\subsection{Leading-order effective Lagrangian}
Let us now turn on $A_\mu$ and the curved metric.  The leading-order
effective Lagrangian, as we have discussed, should be a function of a
finite number of variables which are $O(p^0)$:
\begin{equation}
  \L = \L (\dot\theta, \d_i\theta, A_0, A_i, g_{ij}).
\end{equation}
Gauge invariance and invariance with respect to three-dimensional
general coordinate transformations (with time-independent $\xi^i$)
limit the Lagrangian to be a function of two variables,
\begin{equation}
  \L = \L \left( D_t\theta,\, g^{ij} D_i\theta D_j\theta\right),
\end{equation}
where
\begin{equation}
  D_t\theta = \dot\theta - A_0, \qquad
  D_i\theta = \d_i\theta - A_i \,.
\end{equation}

Now we require the invariance of the effective theory with respect to
general coordinate transformations
\begin{equation}\label{nonrel-gci-theta}
\begin{array}{ll}
  \delta\theta = -\alpha -\xi^k \d_k\theta, & 
    \delta A_0 = -\dot\alpha-\xi^k\d_k A_0 - A_k \dot\xi^k ,\\
   \delta g_{ij} = -\xi^k\d_k g_{ij} -g_{ik}\d_j\xi^k -
  g_{kj}\d_i\xi^k , \qquad &
  \delta A_i = -\d_i\alpha-\xi^k\d_k A_i - A_k\d_i\xi^k +
  mg_{ik}\dot\xi^k .
\end{array}
\end{equation}
Under~(\ref{nonrel-gci-theta}) we have
\begin{align}
  \delta(D_t\theta) &= -\xi^k \d_k D_t\theta - \dot\xi^k D_k \theta ,\\
  \delta(g^{ij} D_i\theta D_j\theta) &=
  -\xi^k \d_k (g^{ij} D_i\theta D_j\theta) - 2m\dot\xi^k D_k\theta .
\end{align}
This means that only one combination of these two variables transforms
as a scalar, which is
\begin{equation}
  X = D_t \theta - \frac{g^{ij}}{2m} D_i\theta D_j\theta,\qquad
  \delta X = -\xi^k \d_k X ,
\end{equation}
and the leading-order effective Lagrangian has to be a function of
this combination,
\begin{equation}\label{LX}
  \L = P(X) .
\end{equation}
If we now set the metric to flat and external fields to zero, then we
find the most general form of the Lagrangian for superfluids,
\begin{equation}\label{GWW-L}
  \L = P \left(\dot\theta - \frac{(\d_i\theta)^2}{2m}\right),
\end{equation}
which was previously found by Greiter, Wilczek, and
Witten~\cite{Greiter:1989qb} using a different line of arguments.

\subsubsection*{Lagrangian density and thermodynamic pressure}

We now show that the function $P$ in the leading-order Lagrangian is
identical to the function that defines, at zero temperature, the
pressure as a function of the chemical potential, $P = P(\mu)$.  For
that purpose, we notice that the particle number density can be found
by differentiating the Lagrangian with respect to $A_0$, and is equal
to
\begin{equation}
  n = P'(X).
\end{equation}
The ground state at a chemical potential $\mu$ corresponds to
$\theta=\mu t$, which means $X=\mu$.  This implies $P$ is a function
satisfying $n(\mu)=P'(\mu)$ where $n(\mu)$ is the equilibrium number
density at chemical potential $\mu$.  That means $P$ is the
thermodynamic pressure, up to a constant.  The constant is not
important in any situation and will be set to zero.

\subsubsection*{Nonrelativistic superfluid as a limit of relativistic 
superfluid}

It is also possible to obtain the Lagrangian~(\ref{GWW-L}) by taking
the nonrelativistic limit of the effective Lagrangian of relativistic
superfluids.  The field appearing in the effective Lagrangian is the
phase $\Theta$ of the relativistic field $\Psi$ (see Eq.~(\ref{relat-S}),
\begin{equation}
  \Psi = |\Psi| e^{-i\Theta}.
\end{equation}
From Eq.~(\ref{Psipsi}), the relation between $\Theta$ and the
nonrelativistic phase $\theta$ is
\begin{equation}\label{Phiphi}
  \Theta = mc^2t + \theta.
\end{equation}
To leading order, the effective Lagrangian is a function of
$\d_\mu\Theta$.  Due to Lorentz invariance, it has to have the form
\begin{equation}
  \L = \L_0(T), \qquad T = - \frac12 g^{\mu\nu}\d_\mu\Theta \d_\nu\Theta.
\end{equation}
From Eqs.~(\ref{Phiphi}) and (\ref{Ginv}) one finds
\begin{equation}
  T = \frac{m^2c^2}2 + m\left[\dot\theta-A_0 - \frac{g^{ij}}{2m}
  (\d_i\theta-A_i)(\d_j\theta-A_j)\right]
  + O(c^{-2})
  = \frac{m^2c^2}2 + mX +O(c^{-2}).
\end{equation}
So, in the nonrelativistic limit the Lagrangian is a function of $X$,
as found in Eq.~(\ref{LX}).

It is shown in Ref.~\cite{Son:2002zn} that the function $\L_0(T)$
coincides with pressure at chemical potential equal to
$\sqrt{2c^2T}=mc^2+X+O(c^{-2})$.  This is consistent with our previous
result, since the difference between the relativistic and
nonrelativistic normalizations of the chemical potential is the rest
energy $mc^2$.

\subsection{Relation to other approaches}

For completeness, we include here a discussion of the relation of the
Lagrangian~(\ref{GWW-L}) and its derivation with other, more
frequently used, approaches.  We also derive some well-known results
using the effective Lagrangian.

\subsubsection*{General coordinate invariance versus Galilean
invariance}

Galilean invariance is sufficient to obtain the leading order
Lagrangian.  We recall that the Galilean transformations
(\ref{alphaxiGal}) are particular cases of general coordinate
transformations.  The Galilean transformations leave flat metric
unchanged.  Under these transformations,
\begin{equation}
  D_t\theta \to D_t\theta - v_i D_i\theta, \qquad
  D_i\theta \to D_i\theta - mv_i,
\end{equation}
from which we see that the only combination which is rotationally and
Galilean invariant is $D_t\theta - (2m)^{-1}(D_i\theta)^2$.

A natural question arises: does Galilean invariance in general
guarantee general coordinate invariance?  The answer is no.  We shall
see in Sec.~\ref{sec:NLO} that at the next-to-leading order, Galilean
invariance is not sufficient to constrain the form of the effective
Lagrangian.

\subsubsection*{Effective Lagrangian and superfluid hydrodynamics}

From Eqs.~(\ref{GWW-L}) and (\ref{thetavarphi}) we can derive the
field equation for $\varphi$,
\begin{equation}\label{continuity}
  \d_t n + \frac1m \d_i (n\d_i \varphi) =0, \qquad 
\end{equation}
In Eq.~(\ref{continuity}) $n$ is the equilibrium number density at
chemical potential equal to $X=\dot\theta-(\d_i\theta)^2/(2m)$.  This
is the continuity equation of superfluid hydrodynamics, if we identify
the superfluid velocity with the gradient of $\theta$,
\begin{equation}
  \v_s = -\frac{\bfnab\theta}m = \frac{\bfnab\varphi}m\,.
\end{equation}
In superfluid hydrodynamics one regards $n$ and $\varphi$ as
independent variables.  The second equation of superfluid
hydrodynamics describes time evolution of $\varphi$ comes from the
definition of $n$,
\begin{equation}
  \dot\varphi = -\mu(n) - \frac{mv_s^2}2\,.
\end{equation}
Here $\mu(n)$ is the chemical potential at which the number density is
equal to $n$.

\subsubsection*{Phonon velocity and interaction} 

We can determine the phonon spectrum by expanding the
Lagrangian~(\ref{GWW-L}) to quadratic order in $\varphi$ (recall that
$\theta=\mu t-\varphi$).  The Lagrangian is then
\begin{equation}
  L = P(\mu) - n\dot\varphi + \frac12\frac{\d n}{\d\mu}\dot\varphi^2
      - \frac n{2m}(\d_i\varphi)^2.
\end{equation}
The first term is constant, and the second term is a full derivative
and does not contribute to the field equation.  The two last terms
determine the phonon speed to be
\begin{equation}
  c_s = \sqrt{\frac nm \frac{\d\mu}{\d n}} = \sqrt{\frac{\d P}{\d\rho}}
  \,.
\end{equation}
In performing the last transformation we have use the thermodynamic
relation $dP=nd\mu$ and the definition of mass density $\rho=mn$.  The
speed of sound is given by the same formula as in normal fluids
(however here the sound is not collisional.)

Expanding the Lagrangian further, one can find the interaction
vertices.  The interaction between phonons, at leading order in
momentum, is determined solely from the equation of state.  This
nontrivial result is actually not new.  Its relativistic equivalence
was derived in Ref.~\cite{Son:2002zn}.  In the nonrelativistic
setting, it was known in the Hamiltonian formalism, which we discuss
next.

\subsubsection*{Effective Hamiltonian}

From the effective Lagrangian~(\ref{GWW-L}) one can derive the
equivalent Hamiltonian.  The variable conjugate to the phase $\theta$
is the number density,
\begin{equation}
  n \equiv \pi_\theta = \frac{\d L}{\d\dot\theta} = P'(X).
\end{equation}
The equal-time commutation relation between $n$ and $\theta$ is
\begin{equation}
  [n(t,\x),\, \theta(t,\y)] = -i \delta(\x-\y).
\end{equation}
The Hamiltonian is then
\begin{equation}
  H = n\dot\theta - P(X) = n \frac{(\nabla\theta)^2}{2m} + (n X -P(X)).
\end{equation}
Note that the function $nX-P(X)$ is the Legendre transform of $P(X)$.
Since $P(X)$ is the pressure as function of chemical potential, its
Legendre transform is energy as function of number density
$\epsilon(n)$.  Therefore,
\begin{equation}
  H = n\frac{(\nabla\theta)^2}{2m} + \epsilon(n) .
\end{equation}
The Hamiltonian is the sum of the kinetic energy $\rho v_s^2/2$ and
the internal energy not associated with the macroscopic flow,
$\epsilon(n)$.  This Hamiltonian is well known.  It was used, e.g., to
find the interaction between phonons~\cite{LL9}.  The Lagrangian
$-n\dot\theta-H$ is sometimes called Popov's hydrodynamic
Lagrangian~\cite{Popov}.

\subsection{Effective Lagrangian for a unitary Fermi gas}

The equation of state of a unitary Fermi gas is known up to a
constant.  Due to the absence of any internal scale, by dimensional
analysis the pressure has to be of the form
\begin{equation}
  P = c_0 m^{3/2}\mu^{5/2},
\end{equation}
where $c_0$ is a dimensionless variable.  The leading-order effective
action is then,
\begin{equation}\label{S-conf}
  S = \int\!dt\,d\x\,c_0 m^{3/2} \left(D_t\theta
      -\frac{g^{ij}}{2m}D_i\theta D_j\theta\right)^{5/2}.
\end{equation}
The fractional power in Eq.~(\ref{S-conf}) is not problematic, since
we always expand around the ground state $\theta=\mu t$.
As expected, the Lagrangian~(\ref{S-conf}) is invariant under 
infinitesimal conformal transformations
\begin{equation}\label{conf-theta}
\begin{array}{ll}
  \delta\theta = -\beta\dot\theta, &
  \delta A_0 = -\beta \dot A_0 -\dot\beta A_0,\\
  \delta g_{ij} = -\beta \dot g_{ij}+\dot \beta g_{ij} , \qquad &
  \delta A_i = -\beta \dot A_i .
\end{array}
\end{equation}

Taking the Legendre transform of $P(\mu)$, we find the energy density
as a function of the number density.  This is proportional to the
energy density of a free Fermi gas at the same density, with the
coefficient of proportionality $\xi$,
\begin{equation}
  \epsilon = \frac35 \left(\frac{2n}{5c_0}\right)^{2/3} \frac nm
  = \xi \frac35 \frac{(3\pi^2 n)^{2/3}}{2m} n.
\end{equation}
The parameter $\xi$ is estimated to be around
$0.4$ from quantum Monte Carlo simulations~\cite{Carlson,Giorgini}
(but see also Ref.~\cite{Lee:2005it}).  We can
express $c_0$ via $\xi$,
\begin{equation}\label{c0xi}
  c_0 = \frac{2^{5/2}}{15\pi^2\xi^{3/2}}\,.
\end{equation}

At large, but finite, scattering length $a$ the effective Lagrangian
contains an extra term proportional to $a^{-1}$,
\begin{equation}\label{Lcd}
  \L = c_0 m^{3/2} X^{5/2} + d_0 \frac{m X^2}a
\end{equation}
where $d_0$ is some constant.  The term proportional to $d_0$ violates
scale and conformal symmetries explicitly.

\section{Superfluid EFT to NLO}
\label{sec:NLO}

\subsection{NLO terms versus loops}

The leading-order effective Lagrangian allows one to compute all
quantities of interest, e.g., scatterings amplitudes or correlation
functions, to leading order in momentum.  The corrections to these
leading-order results come from two sources.  First, the Lagrangian
contains terms subleading in derivatives; these terms contribute, at
tree level, subleading contributions to the scattering amplitudes and
correlation function.  As explained above, these corrections start at
$O(p^2)$ order.  The second source of corrections are loop diagrams,
constructed using the leading-order Lagrangian.

We shall demonstrate here that the loop contributions start at order
$O(p^4)$, i.e., are comparable to those coming from the
next-to-next-to-leading order terms in the effective Lagrangian.
Therefore, if one is interested in the leading $O(1)$ and
next-to-leading $O(p^2)$ results, one can work at tree level.

To estimate the size of loop corrections, it is useful to work in the
unit system where the sound speed is $c_s=1$, so energy and momentum
has the same dimension.  The dimension of the Lagrangian is 4
(momentum$^4$).  We also rescale $\varphi$ so that the quadratic terms
in the Lagrangian is normalized canonically.  The dimension of
$\varphi$ is then 1.  The interaction terms will have coefficients
which can be estimated from dimensional analysis,
\begin{equation}
  \L \sim (\d\varphi)^2 + \frac{O(1)}{\mu^2} (\d_0\varphi)(\d_i\varphi)^2
         + \frac{O(1)}{\mu^4} (\d\varphi)^4.
\end{equation}
Here $\mu$ is the chemical potential and all dimensionless
coefficients are, in general, of order one. 

\begin{figure}[ht]
\centerline{\epsfig{file=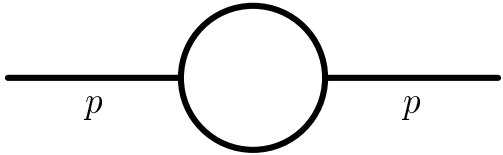,width=0.4\textwidth}}
  \caption[]{A one loop contribution to the phonon self-energy.}
  \label{fig:SelfE}
\end{figure}

As an example of a loop diagram, consider a one-loop correction to the
phonon dispersion relation (Fig.~\ref{fig:SelfE}).  The graph is
divergent and has to be regularized.  After dimensional regularization
it becomes a finite function of the external momentum $p$, with a
possible logarithmic singularity.  However we know that the graph
contains two three-phonon vertices and hence is proportional to
$\mu^{-4}$.  The self-energy has dimension 2.  Therefore we conclude
that
\begin{equation}
  \Sigma_{\rm 1-loop} \sim \frac{p^4}{\mu^4}p^2 \ln p.
\end{equation}
which is suppressed by $O(p^4\ln p/\mu^4)$ compared to the tree-level result
$O(p^2)$.

\begin{figure}[ht]
\centerline{\epsfig{file=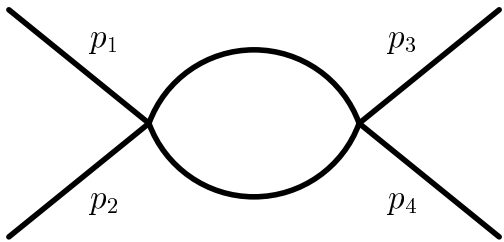,width=0.4\textwidth}}
  \caption[]{A one loop contribution to the phonon scattering amplitude.}
  \label{fig:Scatt}
\end{figure}

Consider another example, which is a one-loop correction to
phonon-phonon scattering amplitude (Fig.~\ref{fig:Scatt}).  The
scattering amplitude is $O(p^4/\mu^4)$ at tree level.  The one-loop
graph contains two four-phonon vertices and hence is proportional to
$\mu^{-8}$.  By dimension, it has to behave as $p^8\ln p/\mu^8$ as a
function of the external momenta $p$.  The result is by a factor of
$O(p^4\ln p/\mu^4)$ smaller than the tree level.

Generalizing to more complicated diagrams and processes, we see that
we have to compute loops only when we want to know the answer to order
$p^4$ compared to the leading-order result.  If we want to know the
answer only to the leading and next-to-leading order, tree-level
calculation using the Lagrangian with subleading terms should suffice.

\subsection{Structures of NLO terms}

To find all terms allowed to the next-to-leading order, it is most
convenient to work in the relativistic theory and then take the
nonrelativistic limit.

In the relativistic theory, the next-to-leading terms contain two
derivatives more than the field $\Theta$.  They have the form
$Q_iF(X)$, $i=1\ldots7$, where $F$ is an arbitrary function of X, and
$Q_i$ are Lorentz scalars which are of order $p^2$:
\begin{equation}\label{list}
\begin{split}
  &Q_1= (\nabla_\mu\d_\nu \Theta) (\nabla_\lambda \d_\rho \Theta) 
  \d^\mu\Theta \d^\nu\Theta \d^\lambda\Theta \d^\rho\Theta 
  = \d_\mu T\d_\lambda T \d^\mu\Theta \d^\lambda\Theta ,\\
  &Q_2=(\nabla_\mu\d_\nu\Theta)(\Box\Theta)\d^\mu\Theta\d^\nu\Theta
  = -\d_\mu T \d^\mu\Theta \Box\Theta,\\
  &Q_3=(\nabla_\mu\d_\nu \Theta) (\nabla^\mu \d_\rho \Theta) 
  \d^\nu\Theta \d^\rho\Theta =\d_\mu T\d^\mu T, \\
  &Q_4=(\Box\Theta)^2, \quad 
  Q_5=(\nabla_\mu\d_\nu\Theta)(\nabla^\mu\d^\nu\Theta), \quad
  Q_6=R^{\mu\nu}\d_\mu\Theta\d_\nu\Theta,\quad
  Q_7=R .
\end{split}
\end{equation}
Here $\nabla_\mu$ is the covariant derivative,
$\Box\Theta\equiv\nabla_\mu\d^\mu\Theta$, $R_{\mu\nu}$ is the Ricci
tensor, and $R\equiv R^\mu_\mu$ (see Ref.~\cite{LL2} and
Appendix~\ref{sec:ricci}).  The terms proportional to the Ricci tensor
are $O(p^2)$ since $R_{\mu\nu}$ contains second derivatives of
$g_{\mu\nu}\sim O(1)$.  

We now use the field equation to exclude redundant terms
from~(\ref{list}).  To leading order, where the Lagrangian is
$\L=\L_0(T)$, the field equation can be written as $\d_\mu T
\d^\mu\Theta = - (\L'/\L'')\Box\Theta$; after adding $O(p^2)$
terms~(\ref{list}) to the Lagrangian, there will be higher corrections
to this equation:
\begin{equation}
  \d_\mu T \d^\mu\Theta = - \frac{\L'}{\L''}\Box\Theta + O(p^3).
\end{equation}
By using the field equation one can express $Q_1$ and $Q_2$ through
$Q_4$:
\begin{equation}
  Q_1 = \left(\frac{\L'}{\L''}\right)^2 Q_4 + O(p^4), \qquad
  Q_2 = \frac{\L'}{\L''} Q_4 + O(p^4),
\end{equation}
so they are not independent.  In addition, one term from~(\ref{list})
can be written as a linear combination of other terms, up to a full
derivative,
\begin{multline}\label{Q5}
  \nabla_\mu \d_\nu\Theta \nabla^\mu \d^\nu\Theta F(T)
  = (\Box\Theta)^2 \left( F- \frac{\L'}{\L''}F'\right) 
    +\d_\mu T \d^\mu T F'(T) - R^{\mu\nu}\d_\mu\Theta \d_\nu\Theta F(T)\\
   - \nabla_\mu\left[(\d^\mu T + \d^\mu\Theta\,\Box\Theta)F(T)
   \right].
\end{multline}
Therefore there are only four independent terms remaining from the
original seven, and
\begin{equation}\label{LNLO-rel}
  \L_{\rm NLO} = 
  \d^\mu T\d_\mu T F_1(T) + (\Box\Theta)^2 F_2(T) 
      + R_{\mu\nu}\d^\mu\Theta\d^\nu\Theta F_3(T) + R F_4(T),
\end{equation}
where $F_1$, $F_2$, $F_3$, and $F_4$ are arbitrary functions. 

Now all we have to do is to take the nonrelativistic limit of the
terms appearing in (\ref{LNLO-rel}).  Consider the $F_1$ term.  Notice
that
\begin{equation}
  \d_0 T = \d_0 \left( \frac{m^2c^2}2 + mX \right) =
  \frac mc \d_t X = O(c^{-1}),
\end{equation}
therefore
\begin{equation}
  \d^\mu T \d_\mu T = m^2 g^{ij}\d_i X \d_j X + O(c^{-2}).
\end{equation}
For the $F_2$ term, we first recall that
\begin{equation}
  \d^\mu \Theta = g^{\mu\nu}\d_\nu\Theta \approx (-mc,\, 
    g^{ij}D_j\theta) ,
\end{equation}
therefore
\begin{equation}
  \Box\Theta = \frac1{\sqrt{-g}}\d_\mu (\sqrt{-g}\,g^{\mu\nu}\d_\nu\Theta)
  = -\frac m{\sqrt g} \d_t \sqrt g 
  + \frac1{\sqrt g} \d_i(g^{ij} D_j\theta).
\end{equation}
The coefficient in front of $F_3$ can be expanded as
\begin{equation}
  R_{\mu\nu} \d^\mu\Theta \d^\nu\Theta 
  = m^2c^2 R_{00} - 2mc R_{0i} D^i\theta + R_{ij} D^i\theta D^j\theta
\end{equation}
where all three terms survive in the limit $c\to\infty$ since $R_{00}=
O(c^{-2})$, $R_{0i}=O(c^{-1})$, and $R_{ij}=O(1)$ [see
Appendix~\ref{sec:ricci} where the components of the Ricci tensor are
computed for the nonrelativistic metric~(\ref{GAg})].

The coefficient in front of $F_4$ is the Ricci scalar, which reduces
to the Ricci scalar of the 3D space in the nonrelativistic limit after
neglecting terms of order $c^{-2}$ and higher:
\begin{equation}
  R = g^{\mu\nu} R_{\mu\nu} \approx g^{ij} R_{ij} = R_{\rm 3D}(g_{ij}).
\end{equation}

Thus, we arrive to the most general NLO Lagrangian consistent with
general coordinate invariance,
\begin{equation}
  \L_{\rm NLO} = \L_1 + \L_2 + \L_3 + \L_4,
\end{equation}
where
\begin{align}
  \L_1 &= g^{ij} \d_i X \d_j X f_1(X),\\
  \L_2 &= \frac1g\left[ -m \d_t\sqrt g + 
        \d_i(\sqrt g\,g^{ij} D_j \theta)\right]^2 f_2(X),\\
\begin{split}
  \L_3 & = \biggl\{-\frac{m^2}4 (2 g^{ij}\ddot g_{ij}
     +\dot g^{ij}\dot g_{ij})
           -m \nabla_i E^i + \frac14 F^{ij}F_{ij}\\
  &\qquad + \Big[m\d_t\d_i\ln g -\nabla_j(mg^{jk}\dot g_{ki} +
   g^{jk}F_{ki}) \Big] D^i\theta + R_{ij} D^i\theta D^j\theta
  \biggr\} f_3(X),
\end{split}\\
  \L_4 &= R_{\rm 3D} f_4(X),
\end{align}
where $E_i\equiv\d_tA_i-\d_iA_0$, $F_{ij}\equiv\d_iA_j-\d_jA_i$, and
raising/lowering indices and covariant derivatives are done using the
3D metric $g_{ij}$.

Now having made use of the general coordinate invariance, we dispense
with the curved space and set the metric to flat,
$g_{ij}=\delta_{ij}$.  The NLO Lagrangian becomes much simpler,
\begin{equation}\label{LNLO-flat}
  \L_{\rm NLO} = \d_i X\d_i X f_1(X) + (\d_i D_i\theta)^2 f_2(X) +
  \left(-m \d_i E_i + \frac14 F_{ij}F_{ij} 
  -  \d_i F_{ij} D_j\theta\right)f_3(X).
\end{equation}
The $f_4$ term is not important in flat space.  If one retains only
$A_0$ and sets $A_i=0$ (which would preclude one from calculating the
transverse response function, see Sec.~\ref{sec:applications}), then
the Lagrangian is even simpler
\begin{equation}\label{LNLO-trap}
  \L_{\rm NLO} = \d_i X\d_i X f_1(X) + (\nabla^2\theta)^2 f_2(X) +
   m \nabla^2 A_0 f_3(X).
\end{equation}

\subsubsection*{General coordinate versus Galilean invariance}

We have mentioned above that, at the leading order, general coordinate
invariance is not more powerful than Galilean invariance: both yield
the Greiter-Wilczek-Witten Lagrangian.  However, at the next to
leading order, general coordinate invariance is more restrictive than
Galilean invariance.  In other worlds, one can write down Galilean
invariant terms which are not general coordinate invariant.

To see that, we write down the infinitesimal Galilean transforms
(recall that we are in flat space),
\begin{equation}
\begin{array}{ll}
  D_t\theta \to D_t\theta - v_i D_i\theta,\qquad & E_i \to E_i - v_k F_{ki},\\
  D_i\theta \to D_i\theta - mv_i, & F_{ij} \to F_{ij}.
\end{array}
\end{equation}
All three terms in Eq.~(\ref{LNLO-flat}) are Galilean invariant, as
they should be since general coordinate invariance implies Galilean
invariance.  However, one can directly check that each of the
following three expressions is Galilean invariant,
\begin{equation}
  F_{ij} F_{ij}, \quad
  m \d_i E_i + \d_i F_{ij} D_j\theta, \quad
  m^2 E_i^2 + 2m E_i F_{ik} D_k\theta 
     + F_{ij} F_{ik} D_j\theta D_k\theta,
\end{equation}
but, in fact, the third expression does not enter the
Lagrangian~(\ref{LNLO-flat}) at all, and the first two enter the
Lagrangian only through one particular linear combination.  Therefore
if one relies only on Galilean invariance one would conclude
incorrectly that the NLO Lagrangian involves five arbitrary functions,
instead of three.  When one only keeps the trapping potential $A_0$,
setting $A_i=0$, general coordinate invariance allows one to exclude a
term of the form $(\d_i A_0)^2f(X)$ from appearing in the
Lagrangian~(\ref{LNLO-trap}).

\subsection{Scale invariance and conformal invariance}

Unitary Fermi gas does not possess an intrinsic scale, therefore the
functions $f_i(X)$, $i=1,\cdots4$ above should be powers of $X$.
Requiring the invariance of the action under scale transformations
$t\to\lambda^{-1} t$, $g_{ij}\to \lambda^{-1}g_{ij}$, one finds that
\begin{equation}
  f_1(X) = c_1 m^{1/2} X^{-1/2}, \quad 
  f_2(X) = c_2 m^{-1/2} X^{1/2}, \quad
  f_3(X) = c_3 m^{-1/2} X^{1/2}.
\end{equation}
The powers of $X$ can be found from counting the scaling dimension and
requiring that $[\L]=\frac52$ (note that $[X]=1$); the powers of $m$
are written to make $c_i$ dimensionless.  Therefore, instead of three
arbitrary functions of $X$, the NLO Lagrangian for unitary Fermi gas
involves three dimensionless phenomenological \emph{constants} $c_1$,
$c_2$, and $c_3$ (four in curved space, with $f_4(X)=c_4
m^{1/2}X^{3/2}$).

Remarkably, conformal invariance further reduces the number of
independent constants to two (three in curved space).  Performing the
infinitesimal transformation~(\ref{conf-theta}), we find that
$S_1=\int\!\L_1$ and $S_4=\int\!\L_4$ are separately invariant.
Indeed, their expressions contain only spatial derivatives of
elementary fields, and thus have conformal dimension equal to the
scaling dimension.  However, $\L_2$ and $\L_3$ contain time
derivatives of the metric tensor and hence do not have well-defined
conformal dimension.  Thus $S_2=\int\!\L_2$ and $S_3=\int\!L_3$ change
under conformal transformations,
\begin{align}
  \delta S_2 &= 3c_2 m^{1/2}\!\int\!dt\,d\x\, \ddot\beta \left[
  m \d_t\sqrt g\, - 
  \d_i(\sqrt g\, g^{ij} D_j\theta)\right] X^{1/2}, \\
  \delta S_3 &= -\frac{c_3}2 m^{3/2}
     \!\int\!dt\,d\x\,\sqrt g\, \dddot\beta X^{1/2}.
  \label{deltaS3}
\end{align}
A closer investigation reveals that the two contributions cancel each
other for a particular ratio of $c_2$ and $c_3$.  To see that, we
perform integration by part in Eq.~(\ref{deltaS3}), and make use of
the leading-order field equation,
\begin{equation}\label{LOeq}
  \dot X = -\frac23 \frac{\d_t\sqrt g}{\sqrt g} X 
  + \frac2{3m} \frac1{\sqrt g}\d_i(\sqrt g\, g^{ij}D_j\theta) X 
  + \frac1m g^{ij} \d_i X \d_j\theta, 
\end{equation}
to obtain
\begin{equation}
  \delta S_3 = c_3m^{3/2}\int\!dt\,d\x\, \ddot\beta \left[
  \frac13 \d_t\sqrt g\, X^{1/2}
  + \frac1{6m} \d_i(\sqrt g\, g^{ij}D_j\theta)X^{1/2}
  + \frac1{4m} \sqrt g\,g^{ij}\d_i X\d_j\theta X^{-1/2}\right].
\end{equation}
The contribution coming from the last term in the brackets can be
further integrated by parts,
\begin{equation}
  \int\!d\x\,\sqrt g\,g^{ij} \d_i X \d_j\theta X^{-1/2} =
  2\!\int\!d\x\, \sqrt g\, g^{ij} \d_i X^{1/2} \d_j\theta =
  - 2\! \int\!d\x\, \d_i(\sqrt g\, g^{ij} \d_j\theta)X^{1/2}.
\end{equation}
One thus finds
\begin{equation}
  \delta S_2 + \delta S_3 = \left(3c_2 + \frac{c_3}3\right)m^{1/2}
  \int\!dt\,d\x\, \ddot\beta 
  [m\d_t \sqrt g- \d_i(\sqrt g\,g^{ij}D_j\theta)] X^{1/2}.
\end{equation}
Therefore, conformal symmetry can be satisfied if one requires
\begin{equation}\label{c3c2}
  c_3 = -9 c_2.
\end{equation}
(In $d$ spatial dimensions we would have $c_3=-d^2 c_2$.)

Thus, we find that the NLO effective Lagrangian of unitary Fermi gas
in flat space and in the presence of background gauge fields involves
only two independent additional constants,
\begin{multline}\label{LA0Ai}
  \L_{\rm LO} + L_{\rm NLO} = 
  c_0 m^{3/2} X^{5/2} + c_1\sqrt m\, X^{-1/2}\d_i X\d_i X\\
  + \frac{c_2}{\sqrt m} 
  \left[(\d_i D_i\theta)^2 + 9m\d_i E_i -\frac94 F_{ij}F_{ij}
  + 9\d_iF_{ij} D_j\theta\right] X^{1/2},
\end{multline}
where $X=\mu-D_t\varphi-(D_i\varphi)^2/(2m)$.  

If we retain only the trapping potential $A_0$ and set $A_i=0$, then
the effective Lagrangian becomes
\begin{equation}\label{LA0}
  \L = 
  c_0 m^{3/2} X^{5/2} + c_1 \sqrt m\,X^{-1/2}\d_i X\d_i X
  + \frac{c_2}{\sqrt m} 
  [(\d_i D_i\theta)^2 - 9m\nabla^2 A_0] X^{1/2}.
\end{equation}

The two parameters $c_1$ and $c_2$ are universal constants and need to
be determined from microscopic computations.  In
Sec.~\ref{sec:applications} we shall see how these coefficients appear
in physical quantities, which suggests methods to extract them.  We
will also show that $c_2$ is positive.

\section{Physical applications}
\label{sec:applications}

In this section, we compute some physical quantities using the NLO
effective Lagrangian.  Our aim is not to go through an exhaustive list
of applications, but just to demonstrate the predictive powers of the
EFT.
Further calculations are deferred to future work.  In this Section we use
the unit system $\hbar=m=1$.

\subsection{Phonon dispersion relation}

Consider an infinite system with no trapping potential $A_0=A_i=0$.
Expanding the effective Lagrangian to quadratic order in $\varphi$, we
find
\begin{equation}
  \L = \frac1{\sqrt\mu}\left[ \frac{15c_0}8\mu \dot\varphi^2 
      - \frac{5c_0}4\mu^2 (\nabla\varphi)^2
      + c_1(\d_i\dot\varphi)^2 
      + c_2\mu (\nabla^2\varphi)^2 \right].
\end{equation}
The phonon dispersion relation is found by solving the equation
\begin{equation}
  \frac{15c_0}8\mu \omega^2 - \frac{5c_0}4\mu^2 q^2
  + c_1\omega^2 q^2 + c_2 \mu q^4 =0.
\end{equation}
The solution is
\begin{equation}\label{omega2q2}
  \omega^2 = \frac{2\mu}3 q^2 - \frac8{45c_0} (2c_1+3c_2) q^4.
\end{equation}
Using $\mu=\xi\epsF$ and Eq.~(\ref{c0xi}), we can write the result in
the form
\begin{equation}
  \omega = \sqrt{\frac\xi3}\, \vF q\left[1
   -\pi^2\sqrt{2\xi}\left(c_1+\frac32c_2\right)\frac{q^2}{\kF^2}\right]. 
\end{equation}
We reproduce the linear part of the phonon dispersion with the
velocity $c_s=\sqrt{\xi/3}\,\vF$, but we also find the nonlinear $q^3$
term in the dispersion.  This correction depends on the combination
$2c_1+3c_2$.  In particular, if $2c_1+3c_2<0$ the dispersion curves up
and a low-energy phonon can decay into two phonons.  In the opposite
case such a process is kinematically forbidden.


\subsection{Static density response function}

The EFT allows one to compute any correlation functions in the
long-wavelength, low-frequency regime.  Here we concentrate on the
static response functions, as these functions are most readily
computable using Monte-Carlo techniques.  The static density response
function $\chi(k)$ is defined through the response of the medium to a
small potential $A_0$~\cite{PinesNozieres}.  If one introduces the
time-ordered density-density correlation function,
\begin{equation}
  i G_{nn}(\omega,q) = \int\!dt\,d\x\, e^{i\omega t-i\q\cdot\x}
    \< T n(t,\x) n(0, \mathbf{0}) \> ,
\end{equation}
then
\begin{equation}
  \chi(q) = G_{nn}(0,q).
\end{equation}
It is related to the dynamic structure function $S(\omega,q)$,
\begin{equation}
  S(\omega,q) = -2\, \textrm{Im}\,G_{nn}(\omega,q),
  \qquad \omega > 0,
\end{equation}
through a dispersion relation
\begin{equation}
  \chi(q) = -2 \int\limits_0^\infty\!\frac{d\omega}{2\pi}\, 
  \frac{S(\omega,q)}\omega \,.
\end{equation}
At zero momentum the static response function is fixed by the
compressibility sum rule,
\begin{equation}
  \chi(0) = -\frac{\d n}{\d\mu} = -\frac n{mc_s^2}\,.
\end{equation}
We shall now compute the leading correction at small $q$.

From the general formalism, the density correlation function can be
obtained by differentiating the generating functional $W$,
\begin{equation}\label{GnnW}
  G_{nn}(t,\x) = -\frac{\delta^2 W[A_0,A_i]}
  {\delta A_0(t,\x)\delta A_0(0,\mathbf{0})}\bigg|_{A_0=A_i=0}.
\end{equation}
In EFT $W$ can be found from diagrammatic techniques.  Since to the
NLO we have only tree graphs, $W$ coincides with the classical action
at the saddle point,
\begin{equation}
  W[A_0, A_i] = S[\varphi_{\rm cl}, A_0, A_i],
\end{equation}
where $\varphi_{\rm cl}$ is the solution to the field equation in the
background $A_0$, $A_i$.

To compute the static density response function one can keep $A_i=0$
and turn on only a static $A_0$.  The leading-order field
equation~(\ref{LOeq}) has a trivial solution $\varphi_{\rm cl}=0$ on
this background.  If one is interested in the NLO action one can just
insert this solution into the NLO Lagrangian.  The Lagrangian density
for this solution is
\begin{equation}
  \L[\varphi_{\rm cl}, A_0] = c_0 (\mu-A_0)^{5/2} +
  \frac{c_1}{\sqrt{\mu-A_0}} (\d_i A_0)^2 
  - 9c_2 (\nabla^2 A_0)\sqrt{\mu-A_0}\,,
\end{equation}
from which we find 
\begin{equation}\label{chic1c2}
  \chi(q) = - \frac{\kF}{\pi^2\xi} \left[
    1 + 2\pi^2\sqrt{2\xi}\left(c_1 - \frac92 c_2\right) 
    \frac{q^2}{\kF^2}\right].
\end{equation}
At $q=0$ the compressibility sum rule is reproduced.  Note that the
$q^2$ correction is proportional to a different linear combination of
$c_1$ and $c_2$ compared to the phonon dispersion relation.

\subsection{Static transverse response function and the sign of $c_2$}

The transverse response function measures the ``paramagnetic'' current
$\j_P$ induced by a transverse static background field ${\bf A}$:
\begin{equation}
  \<\j_\perp(\q)\> = \chi^T(q) {\bf A}_\perp(\q), \qquad
  \q\cdot\j_\perp = \q\cdot {\bf A}_\perp = 0.
\end{equation}
The ``paramagnetic'' current $\j$ in this formula is defined without the
``diagmagnetic part'' ${\bf A}\psi^\+\psi$,
\begin{equation}
  \j = -\frac i2(\psi^\+\bfnab\psi - \bfnab\psi^\+\psi),
\end{equation}
therefore if we define the transverse current-current correlator,
\begin{equation}
  i\delta^{ik}
  G_\perp(\omega,q) = \int\!dt\,d\x\, e^{i\omega t-iqz} 
     \<T j^i(t,\x) j^k(0,{\bf 0})\>, \quad i,k=x,y,
\end{equation}
then the transverse response function is related to $G_\perp$ in the
following way:
\begin{equation}
  \chi^T(q) = G_\perp(0,q) - n.
\end{equation}
At zero temperature, the static transverse response function
approaches zero in the low momentum limit
\begin{equation}
  \lim_{q\to0}\chi^T(q) = 0,
\end{equation}
a fact that corresponds to the vanishing of the normal density.
There is a dispersion relation relating $\chi^T$ with the dynamic
transverse structure function $\Upsilon^T(\omega,q)$:
\begin{equation}
  \chi^T(q) = -2 \int\limits_0^\infty\!\frac{d\omega}{2\pi}\, 
  \frac{\Upsilon^T(\omega,q)}\omega\,,
\end{equation}
where $\Upsilon^T$ is positive, so $\chi^T(q)<0$.

The transverse response function can be computed from a formula
analogous to~(\ref{GnnW}), where $A_0$ is replaced by the ${\bf
A}_\perp$.  Again in the presence of transverse static $A_i$ the field
equation yields $\varphi=0$, and the Lagrangian density is simply
\begin{equation}
  \L[A_i]= c_0 \left(\mu-\frac{\mathbf{A}_\perp^2}2\right)^{5/2} 
  + c_2 \sqrt\mu\left(
     -\frac94 F_{ij} F_{ij} -9 \d_i F_{ij} A_j\right) + 
  O(A_\perp^4). 
\end{equation}
We find
\begin{equation}
  \chi^T(q) = -9c_2 \sqrt{\frac\xi2}\, \kF q^2.
\end{equation}
But since $\chi^T(q)<0$ we conclude that $c_2>0$.

\subsection{Energy of unitary Fermi gas in harmonic trap}

Consider unitary Fermi gas in an asymmetric harmonic trap with the
potential
\begin{equation}
  A_0 = \frac12(\omega_1^2 x_1^2 + \omega_2^2 x_2^2 
        + \omega_3^2 x_3^2).
\end{equation}
To leading order, the ground state at chemical potential $\mu$ is
$\theta=\mu t$, and the Lagrangian density, evaluated for this
configuration and integrated over space, is the free energy
\begin{equation}\label{Ftrap}
  -F(\mu) = \int\d\x\, \left( c_0 (\mu-A_0)^{5/2} 
  + \frac{c_1(\d_i A_0)^2}{\sqrt{\mu-A_0}} 
  - 9 c_2 \nabla^2 A_0\sqrt{\mu -A_0}
  \right).
\end{equation}
In Eq.~(\ref{Ftrap}) the integral is taken over the volume occupied by
the fermion cloud, which is determined by $A_0<\mu$.  Taking the
integrals one obtains
\begin{equation}
  -F(\mu)= \frac1{12\xi^{3/2}} \frac{\mu^4}{\bar\omega^3} +
  \sqrt2\,\pi^2\left(c_1-\frac92c_2\right)
  \frac{\omega_1^2+\omega_2^2+\omega_3^2}{\bar\omega^3}\mu^2,
\end{equation}
where $\bar\omega\equiv(\omega_1\omega_2\omega_3)^{1/3}$.  The
particle number as a function of the chemical potential is found by
differentiating $F(\mu)$, $N = -\d F/\d\mu$, and the energy as a
function of particle number is the Legendre transform of $-F$:
\begin{equation}\label{EN}
  E(N) = F + \mu N = \frac{\sqrt\xi}4 \bar\omega (3N)^{4/3} -
  \sqrt2\, \pi^2\xi \left(c_1 - \frac92c_2\right)
  \frac{\omega_1^2+\omega_2^2+\omega_3^2}{\bar\omega}(3N)^{2/3}.
\end{equation}
Note that $c_1$ and $c_2$ appear in the same combination as they
appear in the static density response function~(\ref{chic1c2}).

We were somewhat fortunate that all integrals in Eq.~(\ref{Ftrap})
converge.  We would not be so lucky beyond the NLO.  Indeed, consider
as an example a $O(q^4)$ term in the effective Lagrangian,
\begin{equation}
  \frac{(g^{ij}\d_i X \d_j X)^2}{X^{7/2}}\,.
\end{equation}
This term respects both general coordinate and conformal invariance.
The contribution to the free energy in a trap would be
\begin{equation}\label{NNLOint}
  \int\!d\x\, \frac{[(\d_i A_0)^2]^2}{(\mu - A_0)^{7/2}}\,.
\end{equation}
For simplicity let us assuming a spherical trap, $\omega_i=\omega$.
Denote the distance to the boundary of the fermion cloud as $\delta$,
\begin{equation}
  |\x| = \frac{\sqrt{2\mu}}\omega - \delta,
\end{equation}
then the integral in~(\ref{NNLOint}), in the region of small $\delta$,
looks like
\begin{equation}\label{intdelta}
  \int\! d\delta\, x^2\frac{(\omega^2x)^4}{(\omega\sqrt\mu\,\delta)^{7/2}}
  \sim \frac{\mu^{5/4}}{\omega^{3/2}}\int\!\frac{d\delta}{\delta^{7/2}}\,.
\end{equation}
Clearly, the integral diverges near $\delta=0$.

This divergence can be traced to the breakdown of the derivative
expansion near the boundary of the cloud.  Indeed, near the boundary
the effective chemical potential $\mu-A_0$ changes substantially over
distances of order $\delta$.  When $\delta$ is comparable to the Fermi
wavelength, i.e.,
\begin{equation}\label{skin}
  \delta \sim \frac1{(\mu-A_0)^{1/2}} 
  \sim \frac1{(\omega\delta)^{1/2}\mu^{1/4}} \qquad
  \textrm{or} \quad \delta \sim \frac1{(\omega^2\mu)^{1/6}}\,.
\end{equation}
the expansion parameter in the EFT becomes of order one.  In other
words, there is a boundary layer with the thickness determined by
Eq.~(\ref{skin}) where the EFT theory break down.  One can estimate
(but not calculate) the contribution of this region to the free energy
by placing (\ref{skin}) as a cutoff in~(\ref{intdelta}), and find
\begin{equation}\label{Fboundary}
  F_{\rm boundary} \sim \frac{\mu^{5/3}}{\omega^{2/3}}\,.
\end{equation}
The EFT, in principle, cannot give an answer with an error smaller
than $F_{\rm boundary}$.  This uncertainty translates into an
uncertainty of order $N^{5/9}$ in Eq.~(\ref{EN}).  Note that this is
smaller than the NLO correction by only a factor of $N^{-1/9}$.

\section{Conclusion}
\label{sec:conclusion}

In this paper we have discussed general coordinate invariance and
conformal invariance in the nonrelativistic many-body theory.  We also
extract physical consequences of these symmetries for the unitary
Fermi gas. We find that to the leading and next-to-leading orders in
the momentum expansion we need three constants $c_0$ (or $\xi$), $c_1$
and $c_2$ to completely characterize the dynamics of the system and
its response to external probes.  While we have some numerical
estimates for $c_0$ from Monte Carlo measurements of thermodynamics,
we still lack estimates for $c_1$ and $c_2$.  We propose to extract
$c_1$ and $c_2$ by finding the density and transverse response
functions in the low-momentum region using Monte Carlo techniques and
matching the results with Eqs.~(\ref{sum:response}).

The effective Lagrangian which we have derived can be used to compute
many physical quantities.  One interesting application is the
computation of the kinetic coefficients of a unitary Fermi gas at low
temperatures.  The knowledge of these coefficients is important for
computing the damping rate of oscillations of the trapped gas at low
temperatures~\cite{Thomas,Grimm-modes}.

Taking a broader perspective, we expect the general coordinate
invariance to have a wide range of applications, since it is a common
property of all nonrelativistic systems.  In the paper we have
considered only systems with one species of particles, but it is
straightforward to generalize the general coordinate
transformations~(\ref{nonrel-gci}) for systems consisting of many
species of particles.  For example, in the case of a $^4$He--$^3$He
mixture one should introduce two gauge fields, one (say, $A_\mu$)
coupled to the $^4$He current and another ($B_\mu$) to the $^3$He
current.  The general coordinate transformations contain two copies of
Eqs.~(\ref{nonrel-gci-A0}) and (\ref{nonrel-gci-Ai}); in particular we
have
\begin{equation}
  \delta A_i = \cdots + m_4 g_{ik} \dot\xi^k, \qquad
  \delta B_i = \cdots + m_3 g_{ik} \dot\xi^k,
\end{equation}
where $m_4$ and $m_3$ are the masses of the $^4$He and $^3$He atoms,
respectively.

On the other hand, conformal invariance, at least in three spatial
dimensions, seems to be specific to the unitary Fermi gas.  It would
be interesting to explore conformal invariance in lower-dimensional
systems, for example for the anyon gas in two dimensions, which has
been suggested to be a superfluid~\cite{Wilczek:1990ik} and respects
scale invariance~\cite{Jackiw:1992fg,Bergman:1993kq}.

\bigskip

{\bf Acknowledgments}\nopagebreak

We thank Eric Braaten, Hans Hammer, and Misha Stephanov for
discussions, and Roman Jackiw and Peter Stichel for comments.
This work is supported, in part, by DOE grant number
DE-FG02-00ER41132.  D.T.S. is supported, in part, by the Alfred
P.~Sloan Foundation.

\appendix

\section{Christoffel symbols and Ricci tensor in the nonrelativistic
limit}
\label{sec:ricci}

Our convention for the Christoffel symbols and the Ricci tensor is
\begin{align}
  \Gamma^\mu_{\nu\lambda} &= \frac12 g^{\mu\rho}(\d_\nu g_{\rho\lambda}
  + \d_\lambda g_{\rho\nu} - \d_\rho g_{\nu\lambda}),\\
  R_{\mu\nu} &= \d_\lambda \Gamma^\lambda_{\mu\nu}
  -\d_\mu \Gamma^\lambda_{\nu\lambda} 
  +\Gamma^\lambda_{\mu\nu}\Gamma^\rho_{\lambda\rho}
  - \Gamma^\lambda_{\mu\rho}\Gamma^\rho_{\nu\lambda}.
\end{align}
The covariant derivatives of scalars and vectors are given by
\begin{equation}
  \nabla_\mu \phi = \d_\mu\phi,\qquad
  \nabla_\mu \xi^\nu = \d_\mu \xi^\nu + 
     \Gamma^\nu_{\mu\lambda}\xi^\lambda, \qquad
  \nabla_\mu \xi_\nu = \d_\mu \xi_\nu -
     \Gamma^\lambda_{\mu\nu}\xi_\lambda.
\end{equation}
To derive Eq.~(\ref{Q5}) the following equation is used:
\begin{equation}
  \nabla_\mu \nabla_\nu \xi^\nu - \nabla_\nu \nabla_\mu \xi^\nu
  = - R_{\mu\nu}\xi^\nu.
\end{equation}

For the metric of the form~(\ref{GAg}), one finds
\begin{eqnarray}
  &&\Gamma^0_{00} = \frac1{mc^3} \dot A_0 + \frac1{m^2c^3}A^i E_i,\\
  &&\Gamma^i_{00} = -\frac1{mc^2} E^i,\qquad
  \Gamma^0_{0i} = \frac1{mc^2}\d_i A_0-\frac1{2mc^2}A^j
    \left(\dot g_{ji} + \frac1m F_{ji}\right),\\
  &&\Gamma^0_{ij} = \frac1{2mc}(\nabla_i A_j+\nabla_j A_i)
     +\frac1{2c}\dot g_{ij}, \qquad
  \Gamma^i_{0j} = \frac1{2c} g^{ik}\left(\dot g_{kj}+\frac1m
  F_{kj}\right),
\end{eqnarray}
where $E_i$ and $F_{ij}$ are the electric and magnetic fields
constructed from the field potential $(A_0, A_i)$:
\begin{equation}
  E_i = \d_t A_i - \d_i A_0, \qquad F_{ij} = \d_i A_j - \d_j A_i.
\end{equation}
The completely spatial components of $\Gamma^\mu_{\nu\lambda}$
coincides with the Christoffel symbols of the three-dimensional space
with metric $g_{ij}$.

For the Ricci tensor, one has
\begin{align}
  R_{00} &= -\frac1{2c^2} g^{ij}\ddot g_{ij}
     -\frac1{4c^2} \dot g^{ij}\dot g_{ij}
           -\frac1{mc^2} \nabla_i E^i + \frac1{4m^2c^2} F^{ij}F_{ij},\\
  R_{0i} &= -\frac1{2c}\d_t\d_i\ln g +\frac1{2c} \nabla_j
  \left[ g^{jk} \left( \dot g_{ki} + \frac1m F_{ki}\right)\right].
\end{align}
Again, the spatial components $R_{ij}$ coincides with the Ricci tensor
of the three-dimensional space with metric $g_{ij}$.  In particular,
in the limit $c\to\infty$, the 4D Ricci scalar $R$ coincides with the
3D counterpart.

\end{document}